\documentclass[aps,prb,twocolumn,superscriptaddress,showpacs,longbibliography]{revtex4-1}
\usepackage{amsmath}
\usepackage{graphicx}
\usepackage{amsfonts}
\usepackage{verbatim}
\usepackage{amssymb}
\usepackage{xcolor}
\usepackage{dsfont}
\usepackage{amsmath} 
\usepackage{hyperref}
\usepackage{physics}
\usepackage{multirow}
\usepackage{makecell}

\hypersetup{colorlinks = true, urlcolor = blue, linkcolor = blue, citecolor = blue}

\newcommand{\bpm}{\begin{pmatrix}}
\newcommand{\epm}{\end{pmatrix}}

\newcommand{\dg}{^{\dagger}}
\newcommand{\hc}{\rm{H.c.}}

\begin{document}

\title{Rabi and Ramsey oscillations of a Majorana qubit in a quantum dot-superconductor array}
\author{Haining Pan}
\affiliation{Department of Physics and Astronomy, Center for Materials Theory, Rutgers University, Piscataway, NJ 08854 USA}
\author{Sankar Das Sarma}
\affiliation{Condensed Matter Theory Center and Joint Quantum Institute, Department of Physics, University of Maryland, College Park, Maryland 20742, USA}
\author{Chun-Xiao Liu}\email{chunxiaoliu62@gmail.com}
\affiliation{QuTech and Kavli Institute of NanoScience, Delft University of Technology, Delft, The Netherlands}

\date{\today}

\begin{abstract}
The Kitaev chain can be engineered within a quantum dot-superconductor array, hosting Majorana zero modes at fine-tuned sweet spots.
In this work, we propose and simulate the occurrence of Rabi and Ramsey oscillations to feasibly construct a minimal Majorana qubit in the quantum dot setup.
Our real-time results incorporate realistic effects, e.g., charge noise and leakage, reflecting the latest experimental progress.
We demonstrate that Majorana qubits with larger energy gaps exhibit significantly enhanced performance---longer dephasing times, higher quality factors, reduced leakage probabilities, and improved visibilities---compared to those with smaller gaps and with conventional quantum dot-based charge qubits. 
We introduce a method for reading out Majorana qubits via quantum capacitance measurements. 
Our work paves the way for future experiments on realizing Majorana qubits in quantum dot-superconductor arrays.
\end{abstract}

\maketitle

\section{Introduction}
Majorana zero modes are non-Abelian anyonic excitations localized at the defects or edges of a topological superconductor~\cite{Read2000Paired, Kitaev2001Unpaired, Nayak2008Non-Abelian, Alicea2012New, Leijnse2012Introduction, Beenakker2013Search, Stanescu2013Majorana, Jiang2013Non, DasSarma2015Majorana, Elliott2015Colloquium, Sato2016Majorana, Sato2017Topological, Lutchyn2018Majorana, Zhang2019Next, Flensberg2021Engineered, DasSarma2023In}.
Qubits constructed from the Majorana excitations are immune to local noise and are fault-tolerant without active error corrections, offering a pathway to implementing error-resilient topological quantum computing~\cite{Nayak2008Non-Abelian, DasSarma2015Majorana}.
Recently the quantum dot-superconductor array has become a promising candidate for realizing topological Kitaev chains~\cite{Kitaev2001Unpaired} in solid-state physics using a concrete idea proposed a while ago~\cite{Sau2012Realizing}.
An advantage of this quantum-dot-based approach is the intrinsic robustness against the effect of disorder that is ubiquitous in semiconductor-superconductor Majorana platforms~\cite{Pan2020Physical, Ahn2021Estimating, DasSarma2023Spectral, DasSarma2023Density, Taylor2024Machine}.
In addition, utilizing Andreev bound states in a hybrid region as coupler enables precise control over the relative amplitudes of normal and superconducting interactions between quantum dots~\cite{Liu2022Tunable, Bordin2023Tunable, Wang2022Singlet, Wang2023Triplet, Bordin2024Crossed}, thus allowing for fine-tuning of a quantum dot-superconductor array into a sweet spot with optimally protected Majorana zero modes~\cite{Kitaev2001Unpaired, Sau2012Realizing, Leijnse2012Parity}.
Tunnel spectroscopic signatures of Majoranas have been observed in recent experiments on quantum dots using both nanowires~\cite{Dvir2023Realization, Zatelli2023Robust, Bordin2024Signatures} and two-dimensional electrons~\cite{tenHaaf2024Twosite}.

\begin{figure}[ht]
    \centering
    \includegraphics[width = 3.4in]{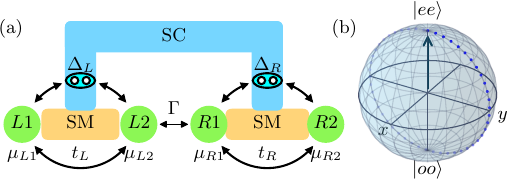}
    \caption{
    (a) Schematic of a Majorana qubit composed of double two-site Kitaev chains. 
    (b) Bloch sphere. $\ket{\pm z}$ are defined as $\ket{ee}$ and $\ket{oo}$, respectively. The dots represent the trajectory of the state vector in a Rabi experiment.
    }
    \label{fig:schematic}
\end{figure}

To decisively establish a Majorana qubit and demonstrate its topologically enhanced coherence, Rabi oscillation experiments on quantum-dot-based Kitaev chains are necessary~\cite{Sau2024Capacitance}.
Additionally, understanding the topological coherence and obtaining a sufficiently long coherence time is crucial for detecting the non-Abelian statistics of Majorana anyons in fusion~\cite{Liu2023Fusion} or braiding experiments~\cite{Boross2024Braiding, Tsintzis2024Majorana}.
Most importantly (and we demonstrate in the current work), such a Rabi oscillation experiment is already feasible in currently available platforms~\cite{Dvir2023Realization, Zatelli2023Robust, tenHaaf2024Twosite, Bordin2024Signatures}, provided that two such minimal Kitaev chains are interconnected via a common superconducting lead, and normal-tunnel-coupled at their ends [see Fig.~\ref{fig:schematic}(a)].

In the current work, we propose Rabi and Ramsey oscillation experiments in a minimal Majorana qubit composed of double two-site Kitaev chains [see Fig.~\ref{fig:schematic}(a)].
Our real-time simulations incorporate realistic effects such as charge noise and leakage to the non-computational bases.
We find that Majorana qubits constructed from large-gap Kitaev chains significantly outperform those with smaller gaps and conventional quantum dot-based charge qubits in terms of dephasing time, quality factor, leakage probability, and visibility.
In addition, we propose a Majorana qubit readout method based on quantum capacitance.
Our work demonstrates the optimal route to the first step of establishing Majorana qubit as a viable experimental entity, which has not been achieved in the fifteen years of experiments~\cite{Mourik2012Signatures, Nichele2017Scaling, Zhang2021Large, Aghaee2023InAs} and twenty-five years of theory~\cite{Read2000Paired, Kitaev2001Unpaired, Nayak2008Non-Abelian, Sau2010Generic, Lutchyn2010Majorana, Oreg2010Helical} on topological quantum computing.

\section{Setup and Hamiltonian}
A minimal Majorana qubit consists of double two-site Kitaev chains, as shown in Fig.~\ref{fig:schematic}(a).
The Hamiltonian is
\begin{equation}\label{eq:H_tot}
    \begin{split}
        & \mathcal{H}_{\text{tot}} = \mathcal{H}_L + \mathcal{H}_R + \mathcal{H}_{\text{tunn}}, \\
        & \mathcal{H}_a = \sum_{i=1}^2\mu_{ai} n_{ai}+ \left( t_a c\dg_{a2} c_{a1} + \Delta_a c_{a2} c_{a1} + \hc \right), \\
        & \mathcal{H}_{\text{tunn}} = \Gamma c\dg_{R1} c_{L2} + \hc 
    \end{split}
\end{equation}
Here $\mathcal{H}_{a}$ with $a\in\left\{ L, R \right\}$ is the Hamiltonian for the left and right chain, respectively, with $\mu_{ai}$ ($i=1,2$) the onsite energy of a spin-polarized dot orbital, $n_{ai}=c_{ai}^\dagger c_{ai}=0,1$ the occupancy number, and $t_a$ and $\Delta_a$ the strengths of the normal and Andreev tunnelings.
$H_{\text{tunn}}$ is the tunnel Hamiltonian, with $\Gamma$ being the strength of single-electron transfer between dots from different chains.
In the current work, we are particularly interested in the sweet spot of the system, which is defined as $\mu_{ai}=0$ and $t_a = \Delta_a$.
{Although $\Delta_L$ and $\Delta_R$ can be different in strength, in the current work we assume them to be equal to simplify the discussions.}
At that point, the even-parity ground state $\ket{e}_a = (\ket{00}_a - \ket{11}_a )/\sqrt{2}$ is degenerate with the odd-parity one $\ket{o}_a = (\ket{10}_a - \ket{01}_a )/\sqrt{2}$ within each Kitaev chain, hosting a pair of Majorana zero modes at two separate quantum dots.
Here $\ket{n_1n_2}_a=\left( c_{a1}^\dagger \right)^{n_1} \left( c_{a2}^\dagger \right)^{n_2}\ket{0}_a$, and $\ket{0}_a$ is the vacuum state of chain-$a$. 
Since total Fermion parity is conserved in the Hamiltonian of Eq.~\eqref{eq:H_tot}, we can focus on the subspace with total parity even without loss of generality. 
As such, the ground-state degeneracy is two-fold:
\begin{align} 
& \ket{ee} \equiv \ket{e}_L \otimes \ket{e}_R, \quad 
\ket{oo} \equiv \ket{o}_L \otimes \ket{o}_R,
\end{align}
which form the basis states of a Majorana qubit.

\begin{figure}[ht]
    \centering
    \includegraphics[width=3.4in]{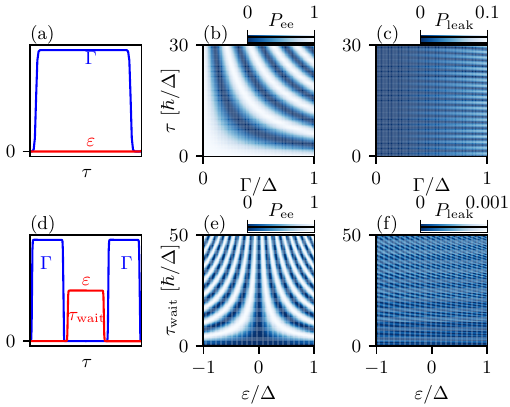}
    \caption{
        Numerical simulations in the clean limit.
        Upper panels: Numerical simulation of a Rabi experiment. (a) Pulse profiles. (b) and (c) $P_{ee}$  and $P_{\text{leak}}$ in Eq.~\eqref{eq:Pleak} in the $(\Gamma, \tau)$ plane.
        Lower panels: Numerical simulation of a Ramsey experiment. (d) Pulse profiles. (e) and (f) $P_{ee}$ and $P_{\text{leak}}$ in the $(\varepsilon, \tau_{\text{wait}})$ plane.
        Here $\Delta_L=\Delta_R=\Delta$.
        }
    \label{fig:Rabi}
\end{figure}

\emph{Rabi oscillations.}---In the qubit subspace spanned by $\ket{ee}$ and $\ket{oo}$, the low-energy effective Hamiltonian is 
\begin{equation}\label{eq:H_eff}
H_{\text{eff}} =  \frac{\varepsilon}{2} \sigma_z + \frac{\Gamma}{2} \sigma_x,
\end{equation}
where $\varepsilon \equiv E_{oo} - E_{ee}$ and $\sigma_{x/z}$ are Pauli X/Z matrices.
Here $\sigma_z$ rotation is proportional to the ground-state energy splitting, which we choose to be $\varepsilon = t_L - \Delta_L$ by detuning the hybrid region in the left chain {away from the sweet spot}~\cite{Zatelli2023Robust}.
$\sigma_x$ rotation is realized by single electron tunneling between the two chains that can be controlled by a tunnel barrier.
Motivated by the form of $H_{\text{eff}}$ in Eq.~\eqref{eq:H_eff}, we perform a numerical simulation of the Rabi and Ramsey experiments using the total Hamiltonian $\mathcal{H}_{\text{tot}}$ in Eq.~\eqref{eq:H_tot}.
Here we implement the qubit rotations by applying sequences of pulses of $\varepsilon$ or $\Gamma$ instead of microwave driving because of the basis state degeneracy. 
In particular, in the Rabi experiment the system is initialized in $\ket{ee}$ of two decoupled Kitaev chains at their sweet spots.
This corresponds to the north pole of the Bloch sphere.
We then turn on the inter-chain tunneling $\Gamma$ and let the system evolve for a time $\tau$ before performing a readout in the $\sigma_z$ basis [see pulse profiles in Fig.~\ref{fig:Rabi}(a)].
Figure~\ref{fig:Rabi}(b) shows the numerically calculated $P_{ee}(\tau) \equiv \abs{\braket{ee}{\psi(\tau)}}^2$ in the $(\Gamma, \tau)$ plane.
Indeed, the fringe pattern of Rabi oscillations confirms that single-electron tunneling $H_{\text{tunn}}$ in Eq.~\eqref{eq:H_tot} works as a $\sigma_x$ rotation in the qubit subspace, with the oscillation frequency being proportional to $\Gamma$.
However, surprisingly, we also find that the state wavefunction can leak out of the qubit subspace with a probability $P_{\text{leak}}(\tau) \equiv 1 - P_{ee}(\tau) -P_{oo}(\tau)$, which oscillates periodically in time and increases with the tunneling strength $\Gamma$ [see Fig.~\ref{fig:Rabi}(c)].
Using time-dependent perturbation theory (see Appendix~\ref{app:leakage}), we show that a finite inter-chain tunneling $\Gamma$ inevitably induces a leakage to the excited states of $\ket{e'e'}$ and $\ket{o'o'}$, i.e.
\begin{align}\label{eq:Pleak}
P_{\text{leak}}(\tau)=P_{e'e'}(\tau)+P_{o'o'}(\tau) 
\approx \frac{\Gamma^{2} }{16 \Delta^{2}} \sin^2 \left( 2\Delta \tau /\hbar \right),
\end{align}
where $\ket{e'}_a = (\ket{00}_a + \ket{11}_a )/\sqrt{2}$ and $\ket{o'}_a = (\ket{10}_a + \ket{01}_a )/\sqrt{2}$ are excited states in each chain and $\Delta_L = \Delta_R=\Delta$.
Here the oscillation frequency of the leakage probability is $4\Delta/\hbar$ and the magnitude scales with $\Gamma^2/\Delta^2$.
On the other hand, in a Ramsey experiment we first apply a pulse of $H_{\text{tunn}}$ to rotate the initial state $\ket{ee}$ to the equator of the Bloch sphere, then let it evolve for a time duration $\tau_{\text{wait}}$ in the presence of a finite $\varepsilon$, and apply the same $H_{\text{tunn}}$ pulse again before the final readout [see pulse profiles in Fig.~\ref{fig:Rabi}(d)].
The simulated $P_{ee}(\tau)$ in the $(\varepsilon, \tau_{\text{wait}})$ plane is shown in Fig.~\ref{fig:Rabi}(e).
Here the small $P_{\text{leak}}$ in Fig.~\ref{fig:Rabi}(f) is due to the $\sigma_x$ pulses, while detuning the coupling $t_L - \Delta_L$ has a negligible impact on the leakage probability.
Both experiments are doable in the currently available devices and provide complementary information about Majorana coherence.

\section{Qubit dephasing}
Charge noise is one of the primary sources of decoherence in semiconductor-based qubits~\cite{Hu2006Charge,Petersson2010Quantum, Dial2013Charge, Scarlino2022Insitu, Connors2022Charge, Throckmorton2022Crosstalk, Burkard2023Semiconductor, Paladino2014f}. 
It can be induced by charge impurities in the environment or fluctuations in the gate voltages nearby.
As a $1/f$ noise, the fluctuations are dominated by the low-frequency components, which can be modeled by the quasi-static disorder approximation, since the zero-frequency part of the noise dominates~\cite{Ithier2005Decoherence, Boross2022Dephasing}.
That is, in each run of the Rabi or Ramsey experiment the Hamiltonian parameters in Eq.~\eqref{eq:H_tot} are subject to a static disorder that obeys normal distribution, and the final readout measurement is averaged over 500 different disorder realizations, giving $\expval{P_{ee}(\tau)}$.
In particular, we simulate and compare three different types of qubits: 1) semiconductor charge qubit with one electron in double quantum dots~\cite{Hayashi2003Coherent, Paladino2014f}, 2) small-gap Majorana qubit~\cite{Dvir2023Realization}, and 3) large-gap Majorana qubit~\cite{Zatelli2023Robust, tenHaaf2024Twosite}.
Here a small (large) gap in the Kitaev chain corresponds to the scenario where the dot-hybrid coupling strength is smaller than (comparable to) the induced gap in the hybrid region~\cite{Liu2024Enhancing}.
The mean values and standard deviations of Hamiltonian parameters that are subject to charge noises are chosen according to the values reported in relevant experimental works, which are summarized in the Appendix~\ref{app:dephasing_mu}.
Figure~\ref{fig:dephasing} shows the calculated Rabi and Ramsey oscillations of $\expval{P_{ee}(\tau)}$ with dephasing for all three types of qubits.
The curves with decaying envelopes are further fitted using the following formula
\begin{equation}\label{eq:Pee}
\expval{P_{ee}(\tau)} = P_0 + A \cos(2\pi f \tau + \phi_0) \exp{ -(\tau/T_2)^{\beta} },
\end{equation}
where $2A$ is the visibility, $T_2$ is the dephasing time, and $\beta$ is the decaying exponent.
Their values are summarized in Table~\ref{tab:qubit_property}, and in addition we define the quality factor as 
\begin{equation}\label{eq:Q}
    Q = 2\pi f T_2,
\end{equation} and the leakage probability as 
\begin{equation}\label{eq:Pleak}
    {P_{\text{leak}}}=\lim_{\tau_0\rightarrow\infty} \int_{0}^{\tau_0} \expval{P_{\text{leak}}(\tau)} d\tau /\tau_0,
\end{equation} 
in the long time limit where $\expval{P_{\text{leak}}(\tau)} = 1 - \expval{P_{ee}(\tau)}-\expval{P_{oo}(\tau)}$ is the instantaneous value.

The Hamiltonian for a semiconductor charge qubit is
\begin{align}
    H_{c} = 
    \bpm
    \varepsilon_L & \Gamma \\
    \Gamma & \varepsilon_R
    \epm,
\end{align}
where the basis states are $\ket{10}$ and $\ket{01}$ with one electron in the left or right quantum dot, $\varepsilon_{L}$ ($\varepsilon_{R}$) is the corresponding orbital energy in the left (right) dot, and $\Gamma$ is the interdot coupling strength.
Here, the fluctuations of the dot energies $\sigma_{\varepsilon}$ dominate the dephasing effect, compared to the fluctuations of the interdot coupling strength $\sigma_{\Gamma}$, due to the small magnitude of $\Gamma$.
In the Rabi experiment, the dot energies are tuned into a sweet spot of $\varepsilon_L=\varepsilon_R=0$, which is insensitive to dot-energy detuning up to the first order, i.e., $\partial E / \partial \varepsilon_a=0$.
However, since the dot-energy fluctuations are large and comparable to the interdot coupling strength, e.g., $\sigma_{\varepsilon}= 3~\mu \text{eV} \simeq \Gamma = 5~\mu$eV, the higher-order contributions (e.g., $\delta E \sim \sigma^2_{\varepsilon}/\Gamma$) lead to a short dephasing time $T_2 \approx 2.50(7)~$ns for $x$ rotations, see Fig.~\ref{fig:dephasing}(a).
In the Ramsey experiment, to implement the $z$ rotation, we choose $\varepsilon_L=-\varepsilon_R=20~\mu$eV $\gg \Gamma$, which is much more susceptible to charge noise as $\partial E / \partial \varepsilon_a \approx 1$.
Thus, the dephasing time is even shorter $T_2 \approx 0.096(2)~$ns and the visibility is reduced, see Fig.~\ref{fig:dephasing}(d).
The consistency between our $T_2$ estimates and the experimental measurements reported in Ref.~\cite{Hayashi2003Coherent} validates our modeling of the quantum dot devices.

In a minimal two-site Kitaev chain that is in the vicinity of the sweet spot, the energy splitting between the even- and odd-parity ground states is approximately $ E \equiv E_{o} - E_{e} \approx \mu_1 \mu_2/2t + (t-\Delta)$, where the first term is due to the simultaneous detuning of onsite dot energies, while the second term is the detuning of the hybrid region.
In a small-gap Majorana qubit (i.e., small $t\equiv\Delta$ limit), the dot-energy fluctuations are comparatively dominant, giving a characteristic energy splitting between the basis states $\delta E \sim \sigma^2_{\mu} / t$.
For a Majorana qubit defined in Fig.~\ref{fig:schematic}(a), such a $\delta E$ leads to noises in the $\sigma_z$ basis.
In the Rabi experiment, since the dot-energy noise ($\propto \sigma_z$) is orthogonal to the $\sigma_x$ rotation, the dephasing effect of the dot energy fluctuations is strongly mitigated (see Appendix~\ref{app:dephasing_mu}).
As such, $T_2 \approx 9.25(5)~$ns is jointly determined by the fluctuations in the dot energies ($\propto \sigma_z$) as well as in the interchain coupling strengths ($\propto \sigma_x$), see Fig.~\ref{fig:dephasing}(b).
In the Ramsey experiment on $ \sigma_z$ rotations, the large dot-energy fluctuations ($\propto \sigma_z$) cause a more detrimental effect on qubit dephasing, giving a much shorter dephasing time $T_2 \approx 1.84(7)~$ns and a reduced visibility $2A \approx 0.3716(8)$, see Fig.~\ref{fig:dephasing}(e) and Table~\ref{tab:qubit_property}.
Note that here the dephasing effect of charge noise in $t_a$ and $\Delta_a$ is negligible because of the weak dot-superconductor hybridization.

On the contrary, the performance of a large-gap Majorana qubit is much improved in almost all aspects, e.g., dephasing time, quality factor, visibility, and leakage probability. 
The strong dot-superconductor hybridization not only strongly enhances the excitation gap of a Kitaev chain, but also transforms the dot orbitals into Yu-Shiba-Rusinov states~\cite{Yu1965Bound,Shiba1968Classical,Rusinov1969Theory}, thus significantly screening the electric charge in the quantum dots~\cite{Liu2024Enhancing, Zatelli2023Robust, tenHaaf2024Twosite}.
As a result, the energy splitting due to $\mu_{ai}$ fluctuations in the effective Kitaev chain is strongly suppressed, i.e., $ \sigma^2_{\mu}/t$ is reduced by a factor of $\sim 300$ compared to the small-gap Majorana qubit.
Now the dominant source of dephasing in the Rabi experiment is the charge noise in $\Gamma$, giving $T_2 \approx 19.064(8)~$ns, see Fig.~\ref{fig:dephasing}(c).
In the Ramsey experiment, the fluctuations of $t_a-\Delta_a$ begin to dominate the dephasing, giving $T_2 \approx 11.23(1)~$ns, see Fig.~\ref{fig:dephasing}(f).
In addition, a larger excitation gap in the Majorana qubit also greatly suppresses the leakage probabilities (see Table~\ref{tab:qubit_property}), consistent with the analytic estimates shown in Eq.~\eqref{eq:Pleak}.

\begin{figure}[ht]
    \centering
    \includegraphics[width=3.4in]{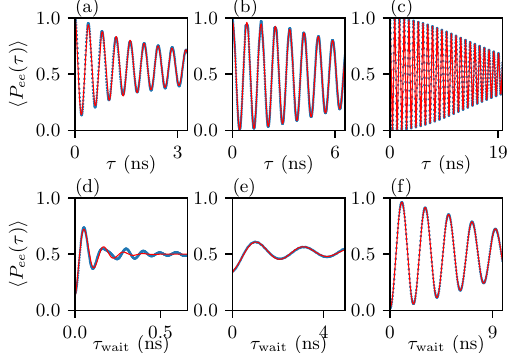}
    \caption{
    Numerical simulations including charge noises.
    Upper panels: Rabi oscillations of disorder-averaged $\expval{P_{ee}}$.
    Lower panels: Ramsey oscillations of disorder-averaged $\expval{P_{ee}}$.
    (a) and (d) semiconductor charge qubits.
    (b) and (e) small-gap Majorana qubits.
    (c) and (f) large-gap Majorana qubits.
    The blue dots are data from numerical simulations, while the red lines are fitting curves using Eq.~\eqref{eq:Pee}.
    Here, the size of the disorder ensemble is 500.
        }
    \label{fig:dephasing}
\end{figure}

\begin{figure}[ht]
    \centering
    \includegraphics[width=3.4in]{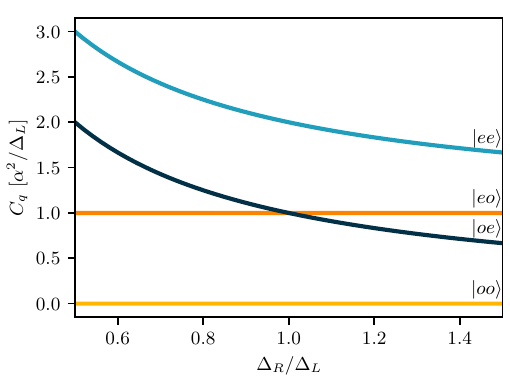}
    \caption{
    Quantum capacitance readout in Eq.~\eqref{eq:Cq} of the low-energy states in a Majorana qubit.
    $\alpha$ is the magnitude of the lever arm of quantum dots, assumed to be identical for all dots.
    $\Delta_{L}$ ($\Delta_{R}$) are the superconducting coupling strengths in the left (right) chains.}
    \label{fig:readout}
\end{figure}

\begin{table}
  \centering
  \caption{Comparison of qubit performances.}
  \label{tab:qubit_property}
  \begin{ruledtabular}
    \begin{tabular}{cccccc}
        \textbf{Protocol} & \makecell{ \textbf{Qubit} \\ \textbf{properties}} & \makecell{\textbf{Charge}\\\textbf{qubit}} & \makecell{\textbf{Small-gap}\\ \textbf{Majorana} \\\textbf{qubit}} & \makecell{\textbf{Large-gap}\\ \textbf{Majorana}\\ \textbf{qubit}}  \\
        \hline
        \multirow{4}{*}{Rabi} & $T_{2}$ (ns) {[Eq.~\eqref{eq:Pee}]}  & 2.50(7) & 9.25(5) & 19.064(8)  \\
        \cline{2-5}
        & $  Q $ {[Eq.~\eqref{eq:Q}]} & 38(1) & 69.4(4) & 144.7(5)  \\
        \cline{2-5}
        & $2A$ {[Eq.~\eqref{eq:Pee}]} & 1.01(2) & 0.958(1) & 1.001(3)  \\
        \cline{2-5}
        & $  P_{\text{leak}} $ {[Eq.~\eqref{eq:Pleak}]} & - & 0.035 & $5.7(8)\times10^{-4}$  \\
        \cline{2-5}
        & $\beta$ {[Eq.~\eqref{eq:Pee}]} & 0.66(2) & 1.87(2) & 1.941(3)  \\
        \hline
        \hline
        \multirow{4}{*}{Ramsey} & $T_{2}$ (ns) {[Eq.~\eqref{eq:Pee}]} & 0.096(2) & 1.84(7) & 11.23(1)  \\
        \cline{2-5}
        & $  Q $ {[Eq.~\eqref{eq:Q}]} & 5.3(2) & 5.6(2) & 34.31(3)  \\
        \cline{2-5}
        & $2A$ {[Eq.~\eqref{eq:Pee}]} & 0.77(2) & 0.3716(8) & 0.950(5)  \\
        \cline{2-5}
        & $  P_{\text{leak}} $ {[Eq.~\eqref{eq:Pleak}]} & - & 0.0275(1) & $3.31(6)\times10^{-5}$  \\
        \cline{2-5}
        & $\beta$ {[Eq.~\eqref{eq:Pee}]} & 1.01(4) & 0.57(2) & 1.557(4)  \\
\end{tabular}
\end{ruledtabular}
\end{table}

\section{Qubit readout}
To read out the Majorana qubits, we consider quantum capacitance measurement as shown in Fig.~\ref{fig:readout}, which is defined as
\begin{align}\label{eq:Cq}
C_q = - \frac{\partial^2 E}{\partial V^2_g}
\end{align}
in the zero-temperature limit~\cite{Liu2023Fusion}.
Here $E$ is the eigenenergy, $V_g$ is the gate voltage that controls the dot energy via $\mu_{ai} = \alpha_{ai} V_g$ with $\alpha_{ai}$ being the lever arm.
Since the measurement is performed when the two chains are decoupled, the result would simply be a sum of the values in each chain, i.e., $C_q = C_{qL} + C_{qR}$.
Furthermore, in the equal-lever-arm regime ($\alpha_{a1} = \alpha_{a2} \equiv \alpha$), the quantum capacitance comes only from the even-parity state within each chain, while that of the odd-parity one is strongly suppressed~\cite{Liu2023Fusion}.
Thus the quantum capacitances of $\ket{ee}$ and $\ket{oo}$ are
\begin{align}
C^{ee}_{q} = \frac{\alpha^2}{ \Delta_L } + \frac{\alpha^2}{ \Delta_R }, \quad C^{oo}_{q} = 0,
\label{eq:C_q}
\end{align}
which are distinct from each other and therefore can be used for qubit readout (see Appendix~\ref{app:readout}).
Following the argument, we further obtain that $C^{eo}_{q} = \alpha^2/ \Delta_L $ and $C^{oe}_{q} = \alpha^2/ \Delta_R $, which are different from both $C^{ee}_{q}$ and $C^{oo}_{q}$.
Therefore in addition to qubit readout, $C_q$ measurement can simultaneously reveal the quasiparticle poisoning effect that transitions a Majorana qubit between states in different total parity space.

\section{Discussion}
In the numerical simulations, we regard $1/f$ charge noise as the dominant source of decoherence in the proposed devices, neglecting the quasiparticle poisoning effect because this is the prevailing situation in semiconductor platforms.
For example, a poisoning time around $\sim1$ ms, as reported in a similar semiconductor-superconductor hybrid device~\cite{Aghaee2024Interferometric}, is much longer than the dephasing time considered here $\sim 10$ ns, making poisoning insignificant for the current consideration where $1/f$ charge noise dominates decoherence.
In addition, here both Rabi and Ramsey experiments are simulated using the most basic protocols for $x$ and $z$ rotations in order to demonstrate the working principles and to provide a fair comparison between semiconductor charge qubits and Majorana qubits.
We emphasize that the system we consider~\cite{Dvir2023Realization, Zatelli2023Robust, tenHaaf2024Twosite, Bordin2024Signatures} is equivalent to semiconductor charge qubits if all superconductivity is removed from considerations.
It is, therefore, possible to further improve the dephasing time, e.g., by optimizing the pulse profiles, by designing a form of interdot coupling that is more resilient against charge noises, or by further scaling up the Kitaev chain~\cite{Bordin2024Crossed, Bordin2024Signatures, Liu2024Protocol}.
Such considerations should be relevant once the basic Rabi and Ramsey oscillations proposed by us are observed so that the elementary concept of a Majorana qubit is established beyond the simplest transport measurements prevalent so far in this subject. 
We emphasize that our work establishes the feasibility of Rabi oscillations in the already existing experimental platforms of Refs.~\cite{Dvir2023Realization, Zatelli2023Robust, tenHaaf2024Twosite, Bordin2024Signatures}.

\section{Summary}
We propose and simulate Rabi and Ramsey oscillation experiments for a minimal Majorana qubit defined in coupled quantum dot-superconductor arrays.
Our realistic calculations predict actual results of such an experiment and demonstrate that the performance of large-gap Majorana qubits significantly surpasses that of small-gap counterparts and traditional conventional charge qubits, although some enhancement over semiconductor charge qubits should already manifest in the small-gap platforms.
Consequently, conducting such experiments is both feasible and promising on currently available Kitaev chain devices, utilizing existing control and measurement technologies.
Our work thus points in the new experimental direction of establishing the quantum dot Majorana systems as a viable platform by showing how to perform a completely new experiment, which has not been suggested before in the context of Majorana systems.
This would provide a crucial step toward the realization of the first Majorana qubit in solid-state systems.
In fact, the observation of stable Rabi oscillations is synonymous with having a qubit, and our work establishes that such a qubit experiment should be successful in the existing Majorana platforms.
{The observation of Rabi oscillations in this platform will establish that a feasible qubit exists here, and may also establish that this qubit has substantially enhanced coherence compared with the corresponding semiconductor quantum dot based charge qubits with no superconductors.}

\section*{Acknowledgements}
We are particularly grateful to Xin Zhang, Francesco Zatelli, F. Setiawan, Michael Wimmer, and Jay D. Sau for useful discussions.
H.P. is supported by US-ONR grant No.~N00014-23-1-2357.
C.-X.L. is supported by a subsidy for top consortia for knowledge and innovation (TKI toeslag).
S.D.S. is supported by the Laboratory for Physical Sciences through the Condensed Matter Theory Center at the University of Maryland.
\bibliography{references_CXL}

\appendix
\onecolumngrid

\section{Model}
    The system of the double two-site Kitaev chain is described by the following Hamiltonian:
    \begin{equation}\label{eq:H}
        \mathcal{H}=\mathcal{H}_{L} + \mathcal{H}_{R} + \mathcal{H}_{T}+\mathcal{H}_{\mu}
    \end{equation}
    where intra-chain coupling in the left chain (Site index 1 and 2) and right chain (Site index 3 and 4) are
    \begin{equation}\label{eq:H_L_R}
        \begin{split}
            \mathcal{H}_{L}= t \sum_{} \left( c_{L2}^\dagger c_{L1} +c_{L1}^\dagger c_{L2} \right)+\Delta \left( c_{L2}c_{L1}+c_{L1}^\dagger c_{L2}^\dagger  \right)\\
            \mathcal{H}_{R}= t \sum_{} \left( c_{R2}^\dagger c_{R1} +c_{R1}^\dagger c_{R2} \right)+\Delta \left( c_{R2}c_{R1}+c_{R1}^\dagger c_{R2}^\dagger  \right),
        \end{split}
    \end{equation}
    the inter-chain hopping is
    \begin{equation}\label{eq:H_T}
        \mathcal{H}_T= \Gamma \left( c_{R1}^\dagger c_{L2}+c_{L2}^\dagger c_{R1} \right),
    \end{equation}
    and the onsite chemical potential is
    \begin{equation}\label{eq:H_mu}
        \mathcal{H}_{\mu}= \sum_{a\in \left\{ L,R \right\}}\sum_{i=1}^2 \mu_i  c_{a,i}^\dagger c_{a,i}.
    \end{equation}
    Up to a particle-hole transformation, we can choose $t>0$, $ \Delta>0$, and $\Gamma>0$. 

    Here, without the tunneling term $\Gamma=0$, the two chains are decoupled, where the sweet spot is achieved when $t=\Delta$, and $\mu_i=0$, leading to the ground state manifold spanned by
    \begin{equation}
        \ket{e}_L = \frac{1}{\sqrt{2}}\left( 1- c_{L1}^\dagger c_{L2}^\dagger \right)\ket{0},
        \ket{o}_L = \frac{1}{\sqrt{2}}\left( c_{L1}^\dagger -c_{L2}^\dagger \right)\ket{0},
    \end{equation}
    \begin{equation}
        \ket{e}_R = \frac{1}{\sqrt{2}}\left( 1- c_{R1}^\dagger c_{R2}^\dagger \right)\ket{0},
        \ket{o}_R = \frac{1}{\sqrt{2}}\left( c_{R1}^\dagger -c_{R2}^\dagger \right)\ket{0},
    \end{equation}
    for the left and right systems, respectively.

    With the tunneling term, the ground state of the two chains can be spanned by the two other single-chain excited states denoted as
    \begin{equation}
            \ket{e'}_L = \frac{1}{\sqrt{2}}\left( 1+c_{L1}^\dagger c_{L2}^\dagger \right)\ket{0},
            \ket{o'}_L = \frac{1}{\sqrt{2}}\left( c_{L1}^\dagger +c_{L2}^\dagger \right)\ket{0},
    \end{equation}
    \begin{equation}
            \ket{e'}_R = \frac{1}{\sqrt{2}}\left( 1+ c_{R1}^\dagger c_{R2}^\dagger \right)\ket{0},
            \ket{o'}_R = \frac{1}{\sqrt{2}}\left( c_{R1}^\dagger +c_{R2}^\dagger \right)\ket{0}.
    \end{equation}
    Therefore, $\ket{e}_L$, $ \ket{e'}_L$, $ \ket{o}_L$, and $ \ket{o'}_L$ ($\ket{e}_R$, $ \ket{e'}_R$, $ \ket{o}_R$, and $ \ket{o'}_R$) form the complete basis for the left (right) chain.

    Without the loss of generality, we choose to work in the even-total-parity, leading to a complete basis of $\ket{ee}\equiv\ket{e}_L\ket{e}_R$,  $ \ket{oo}$, $\ket{e'e'}$,$ \ket{o'o'}$, $ \ket{ee'}$,  $ \ket{oo'}$, $ \ket{e'e}$, and $ \ket{o'o}$.
    With this set of bases, the matrix representation of the sum of Hamiltonian Eq.~\eqref{eq:H_L_R} and Eq.~\eqref{eq:H_T} is
    \begin{equation}\label{eq:H_mat}
        H_L+H_R+H_T=\begin{pmatrix}
            h_{+} & 0\\
            0 & h_{-}
        \end{pmatrix},
    \end{equation}
    where $h_{+}$ and $h_{-}$ are
    \begin{equation}\label{eq:h+}
        h_+=\begin{pmatrix}
            -2\Delta &  -\Gamma/2 & 0 &-\Gamma/2\\
            -\Gamma/2 &  -2t &\Gamma/2 & 0\\
            0 &  \Gamma/2 &2\Delta & \Gamma/2\\
            -\Gamma/2 &  0 & \Gamma/2 & 2t
        \end{pmatrix},
    \end{equation}
    \begin{equation}\label{eq:h-}
        h_-=\begin{pmatrix}
            0 &  \Gamma/2 &0 & \Gamma/2\\
            \Gamma/2 &  0 & -\Gamma/2 &0\\
            0 &  -\Gamma/2 &0 & -\Gamma/2\\
            \Gamma/2 &  0 &-\Gamma/2 & 0
        \end{pmatrix}.
    \end{equation}

    Similarly, the matrix representation of the onsite chemical potential Eq.~\eqref{eq:H_mu} is
    \begin{equation}\label{eq:H_mu_mat}
        H_\mu=\frac{\mu_{1234}}{2}\mathds{1} +\frac{1}{2}\begin{pmatrix}
            0 & h_{\mu} \\
            h_{\mu}^\dagger & 0
        \end{pmatrix},
    \end{equation}
    where 
    \begin{equation}\label{eq:h_mu}
        h_{\mu}= \begin{pmatrix}
            - \mu_{34} & 0 & - \mu_{12} & 0\\0 & \delta\mu_{34} & 0 & \delta\mu_{12}\\- \mu_{12} & 0 & - \mu_{34} & 0\\0 & \delta\mu_{12} & 0 & \delta\mu_{34}
        \end{pmatrix},
    \end{equation}
    with the shorthand notions of $\mu_{1234}\equiv\sum_{i=1}^4\mu_i$ , $\mu_{12}\equiv\mu_1+\mu_2$, $\mu_{34}\equiv\mu_3+\mu_4$, $\delta\mu_{12}\equiv\mu_1-\mu_2$, and $\delta\mu_{34}\equiv\mu_3-\mu_4$.

    \section{Leakage due to $\Delta$, $t$, and $\Gamma$}\label{app:leakage}
    In Eq.~\eqref{eq:H_eff} in the main text, we considered the disorder effect in $\epsilon$ and $\Gamma$ before $\sigma_z$ and $\sigma_x$.
    Here, we will consider their leakage effect separately to understand the leakage that is effectively on $\epsilon$ and $\Gamma$.
    
    We first consider the effect of the disorder only in $\Delta$, $t$, and $\Gamma$, i.e., $\mu_i=0 $, because Eq.~\eqref{eq:H_mat} is block diagonal, and given the initial state being $\ket{ee}$, we only need to consider the subspace of $h_+$, where the Rabi oscillation is between $\ket{ee}$ and $\ket{oo}$, and the leakage states are $\ket{e'e'}$ and $\ket{o'o'}$.

    We use the time-dependent perturbation theory, where Eq.~\eqref{eq:H_mat} is decomposed into the noninteracting part $H_0$ 
    \begin{equation}\label{eq:H0}
        H_0=\begin{pmatrix}
            -2\Delta & -\Gamma/2 & 0 & 0\\
            -\Gamma/2 & -2t & 0 & 0\\
            0 & 0 & 2\Delta & \Gamma/2\\
            0 & 0 & \Gamma/2 & 2t
        \end{pmatrix},
    \end{equation}
    and the perturbation $H_1$ as 
    \begin{equation}\label{eq:H1}
        H_1=\begin{pmatrix}
            0 & 0 & 0 & -\Gamma/2\\
            0 & 0 & \Gamma/2 & 0\\
            0 & \Gamma/2 & 0 & 0\\
            -\Gamma/2 & 0 & 0 & 0
        \end{pmatrix}.
    \end{equation}
    
    Conceptually, the first term $H_0$ in Eq.~\eqref{eq:H0} accounts for the Rabi oscillation between $\ket{ee}$ and $\ket{oo}$ (given the initial state being $\ket{ee}$), and the second term $H_1$ in Eq.~\eqref{eq:H1} leads to $\ket{e'e'}$ and $\ket{o'o'}$.

    The time evolution operator (in the Schr\"odinger picture) is expanded in the Dyson series (truncated at the first order) as
    \begin{equation}
        U(\tau)=e^{-i H_0 \tau}\left( 1-i\int_{0}^{\tau}d\tau_1 H_{1,I}(\tau_1) \right) e^{iH_0 \tau} ,
    \end{equation}
    where $H_{1,I}(\tau_1)$ is $H_1$ in the interacting picture
    \begin{equation}\label{eq:H1I}
        H_{1,I}(\tau_1)= e^{i H_0 \tau} H_{1} e^{-i H_0 \tau} =\sum_{i,j} \matrixelement{i}{H_{1}}{j} e^{i\left( E_i-E_j \right)\tau_1} ,
    \end{equation} 
    and $\ket{i}$ is the eigenvector of $H_0$ with the eigenvalues of $E_i$.

    The spectrum of $H_0$ in Eq.~\eqref{eq:H0} is
    \begin{equation}\label{eq:eigen0}
        \begin{split}
            -E_+:&\quad\begin{pmatrix}
                \cos(\frac{\theta}{2}) & \sin(\frac{\theta}{2}) & 0 & 0 
            \end{pmatrix}^\intercal\\
            -E_-:&\quad\begin{pmatrix}
                -\sin(\frac{\theta}{2}) & \cos(\frac{\theta}{2})  & 0 & 0 
            \end{pmatrix}^\intercal\\
            E_-:&\quad\begin{pmatrix}
                0 & 0 & -\sin(\frac{\theta}{2}) & \cos(\frac{\theta}{2})   
            \end{pmatrix}^\intercal\\
            E_+:&\quad\begin{pmatrix}
                0 & 0 & \cos(\frac{\theta}{2}) & \sin(\frac{\theta}{2})  
            \end{pmatrix}^\intercal\\
        \end{split}
    \end{equation}
    where $E_\pm = \Delta+t\pm\sqrt{(\Delta-t)^2+(\Gamma/2)^2}$, and $\tan\theta=\frac{\Gamma/2}{\Delta-t}$, 
    We substitute the eigenvectors and eigenvalues of $H_0$ into Eq.~\eqref{eq:H1I}, with the initial state 
    \begin{equation}
        \ket{\psi(\tau=0)}=\ket{ee}= \begin{pmatrix}
            1 & 0 & 0 & 0
        \end{pmatrix}^\intercal,
    \end{equation}
    we have the (unnormalized) final state in the Schr\"odinger picture as
    \begin{equation}
        \ket{\psi(\tau)}=U(\tau)\ket{\psi(\tau=0)}=\ket{\psi(\tau)}=\begin{pmatrix}
            e^{i\omega \tau}\left[ \cos(\Omega \tau )  + i\sin( \Omega \tau )\cos\theta\right]\\
            e^{i\omega \tau}i\sin( \Omega \tau )\sin\theta\\
            \frac{\Gamma}{2\omega}\sin(\omega \tau)\sin(\Omega \tau)\sin\theta\\
            \frac{\Gamma}{2\omega}\left[ -\sin(\Omega \tau)\cos\theta+i\cos(\Omega \tau) \right]  \sin(\omega \tau)
        \end{pmatrix},
    \end{equation}
    where $\Omega=\sqrt{(\Delta-t)^2+(\Gamma/2)^2}$ and $\omega=\Delta+t$.

    Therefore, the Rabi oscillation between $\ket{ee}$ and $\ket{oo}$ have the probability densities of
    \begin{equation}
        \begin{split}
            P_{ee}(\tau)&=\abs{\braket{ee}{\psi(\tau)}}^2=\sin^{2}{\left(\Omega \tau \right)} \cos^{2}{\left(\theta \right)} + \cos^{2}{\left(\Omega \tau \right)} \\
            P_{oo}(\tau)&=\abs{\braket{oo}{\psi(\tau)}}^2=\sin^{2}{\left(\theta \right)} \sin^{2}{\left(\Omega \tau \right)}
            ,
        \end{split}
    \end{equation}
    and the leakage is
    \begin{equation}\label{eq:leak}
        P_{\text{leak}}(\tau)=P_{e'e'}(\tau)+P_{o'o'}(\tau)=\abs{\braket{e'e'}{\psi(\tau)}}^2+\abs{\braket{o'o'}{\psi(\tau)}}^2=\frac{\Gamma^{2} \sin^{2}{\left( \omega \tau \right)}}{4 \omega^{2}},
    \end{equation}
    indicating that the leakage frequency is $2\omega=2(\Delta+t)$, which is independent of the $\Gamma$, and consistent with Fig. 2(b) in the main text.

    Specifically, at the sweep spot $\Delta=t$, we have $\theta=\pi/2$ and the probability densities are
    \begin{equation}
        \begin{split}
            P_{ee}(\tau)&=\cos^{2}\left(\Omega \tau \right)\\
            P_{oo}(\tau)&=\sin^{2}\left(\Omega \tau \right)\\
            P_{\text{leak}}(\tau)& = \frac{\Gamma^{2}}{16\Delta^{2}}\sin^{2}\left(2\Delta \tau \right),
        \end{split}
    \end{equation}
    recovering the Rabi frequency $2\Omega=\Gamma$ in Frg. 2(a).

    \section{Leakage due to $\mu_1$}
    To consider the leakage effect in $\mu_1$ (or, equivalently for $\mu_4$ due to the inversion symmetry), we set all other parameters to the sweet spot, including $\Delta=t_0$ and $\mu_2=\mu_3=\mu_4=0$.  
    Following the same time-dependent perturbation theory, the noninteracting part $H^{\mu_1}_0$ is
    \begin{equation}\label{eq:H_mu1}
        H^{\mu_1}_0=\frac{\mu_1}{2} \mathds{1}+ \begin{pmatrix}
            H_0 & 0 \\
            0 & h_-
        \end{pmatrix}
    \end{equation}
    where $H_0$ and $h_-$ are in Eq.~\eqref{eq:H0} and in Eq.~\eqref{eq:h-} (with $\Delta=t$),  
    and the perturbation term $H^{\mu_1}_1$ is
    \begin{equation}
        H^{\mu_1}_1=\begin{pmatrix}
            H_1 & h_{\mu_1}\\
            h_{\mu_1}^\dagger & 0
        \end{pmatrix}
    \end{equation}
    where $H_1$ is in Eq.~\eqref{eq:H1} and $H_{\mu_1}$ is 
    \begin{equation}
        h_{\mu_1}=\frac{1}{2}\begin{pmatrix}
            0 & 0 & -\mu_1 & 0 \\
            0 & 0 & 0 & \mu_1\\
            -\mu_1 & 0 & 0 & 0\\
            0 & \mu_1 & 0 & 0
        \end{pmatrix}.
    \end{equation}
    from Eq.~\eqref{eq:h_mu}.

    Here, the noninteracting term $H^{\mu_1}_0$ has the eigenvalues and eigenvectors as
    \begin{equation}\label{eq:eigen}
        \begin{split}
            - 2 \Delta - \frac{\Gamma}{2} + \frac{\mu_{1}}{2} :&\quad\frac{1}{\sqrt{2}}\begin{pmatrix}
                1 & 1 & 0 & 0 & 0 & 0 & 0 & 0
            \end{pmatrix}^\intercal\\
            - 2 \Delta + \frac{\Gamma}{2} + \frac{\mu_{1}}{2} :&\quad\frac{1}{\sqrt{2}}\begin{pmatrix}
                -1 & 1  & 0 & 0 & 0 & 0 & 0 & 0
            \end{pmatrix}^\intercal\\
            2 \Delta - \frac{\Gamma}{2} + \frac{\mu_{1}}{2} :&\quad\frac{1}{\sqrt{2}}\begin{pmatrix}
                0 & 0 & -1 & 1 & 0 & 0 & 0 & 0
            \end{pmatrix}^\intercal\\
            2 \Delta + \frac{\Gamma}{2} + \frac{\mu_{1}}{2} :&\quad\frac{1}{\sqrt{2}}\begin{pmatrix}
                    0 & 0 & 1 & 1 & 0 & 0 & 0 & 0
                \end{pmatrix}^\intercal\\
            \frac{\mu_{1}}{2}: & \quad \frac{1}{\sqrt{2}}\begin{pmatrix}
                0 & 0 & 0 & 0 & 1 & 0 & 1 & 0
            \end{pmatrix}^\intercal\\
            \frac{\mu_{1}}{2}: & \quad \frac{1}{\sqrt{2}}\begin{pmatrix}
                0 & 0 & 0 & 0 & 0 & -1 & 0 & 1
            \end{pmatrix}^\intercal\\
            - \Gamma + \frac{\mu_{1}}{2}: & \quad \frac{1}{2}\begin{pmatrix}
                0 & 0 & 0 & 0 & 1 & 1 & -1 & 1
            \end{pmatrix}^\intercal\\
            \Gamma + \frac{\mu_{1}}{2}: & \quad \frac{1}{2}\begin{pmatrix}
                0 & 0 & 0 & 0 & -1 & 1 & 1 & 1
            \end{pmatrix}^\intercal\\
        \end{split}
    \end{equation}

    With the initial state being $\ket{ee}= \begin{pmatrix}
        1 & 0 & 0 & 0 & 0 & 0 & 0 & 0
    \end{pmatrix}^\intercal$, the (unnormalized) final state in the Schr\"odinger picture is
    \begin{equation}
        \psi^{\mu_1}(\tau)=\begin{pmatrix}
            e^{i \tau \left(2 \Delta - \frac{\mu_{1}}{2}\right)} \cos{\left(\frac{\tau \Gamma}{2} \right)}\\
            i e^{i \tau \left(2 \Delta - \frac{\mu_{1}}{2}\right)} \sin{\left(\frac{\tau \Gamma}{2} \right)}\\
            \frac{\Gamma}{4 \Delta}e^{- \frac{i \tau \mu_{1}}{2}} \sin{\left(2 \tau \Delta \right)} \sin{\left(\frac{\tau \Gamma}{2} \right)}\\
            \frac{i \Gamma }{4 \Delta}e^{- \frac{i \tau \mu_{1}}{2}} \sin{\left(2 \tau \Delta \right)} \cos{\left(\frac{\tau \Gamma}{2} \right)}\\
            \frac{1}{2} \mu_{1} e^{i \tau \left(\Delta - \frac{\mu_{1}}{2}\right)} \sin{\left(\frac{\tau \Gamma}{2} \right)}\left( \frac{1}{E^{ }_{-}}e^{\frac{i \tau \Gamma}{4}} \sin{\left(\frac{E^{ }_{-} \tau}{2} \right)} - \frac{1}{E^{ }_{+}} e^{- \frac{i \tau \Gamma}{4}} \sin{\left(\frac{E^{ }_{+} \tau}{2} \right)} \right)\\
            \frac{1}{2}\mu_{1} e^{i \tau \left(\Delta - \frac{\mu_{1}}{2}\right)} \cos{\left(\frac{\tau \Gamma}{2} \right)}\left( - \frac{1}{E^{ }_{-}}i e^{\frac{i \tau \Gamma}{4}} \sin{\left(\frac{E^{ }_{-} \tau}{2} \right)} + \frac{1}{E^{ }_{+}} i e^{- \frac{i \tau \Gamma}{4}} \sin{\left(\frac{E^{ }_{+} \tau}{2} \right)} \right)\\
            \frac{1}{2}\mu_{1} e^{i \tau \left(\Delta - \frac{\mu_{1}}{2}\right)} \cos{\left(\frac{\tau \Gamma}{2} \right)} \left( \frac{1}{E^{ }_{-}}i e^{\frac{i \tau \Gamma}{4}} \sin{\left(\frac{E^{ }_{-} \tau}{2} \right)} + \frac{1}{E^{ }_{+}}i e^{- \frac{i \tau \Gamma}{4}} \sin{\left(\frac{E^{ }_{+} \tau}{2} \right)} \right)\\
            \frac{1}{2}\mu_{1} e^{i \tau \left(\Delta - \frac{\mu_{1}}{2}\right)} \sin{\left(\frac{\tau \Gamma}{2} \right)} \left( \frac{1}{E^{ }_{-}}e^{\frac{i \tau \Gamma}{4}} \sin{\left(\frac{E^{ }_{-} \tau}{2} \right)} + \frac{1}{E^{ }_{+}}e^{- \frac{i \tau \Gamma}{4}} \sin{\left(\frac{E^{ }_{+} \tau}{2} \right)}\right)
        \end{pmatrix},
    \end{equation}
    where $E_{\pm}=2\Delta\pm\Gamma/2$.  
    
    Therefore, the leakage to the $\ket{e'e'}$ and $\ket{o'o'}$ is the same
    \begin{equation}
        P_{e'e'}(\tau)+P_{o'o'}(\tau)=\abs{\braket{e'e'}{\psi^{\mu_1}(\tau)}}^2+\abs{\braket{o'o'}{\psi^{\mu_1}(\tau)}}^2=\frac{\Gamma^{2} \sin^{2}{\left(2 \tau \Delta \right)}}{16 \Delta^{2}},
    \end{equation}
    and the leakage to $ \ket{ee'}$,  $ \ket{oo'}$, $ \ket{e'e}$, and $ \ket{o'o}$ is
    \begin{equation}\label{eq:P_leak_mu1}
        \begin{split}
            P_{ee'}(\tau)+P_{oo'}(\tau)+P_{e'e}(\tau)+P_{o'o}(\tau)&=\abs{\braket{ee'}{\psi^{\mu_1}(\tau)}}^2+\abs{\braket{oo'}{\psi^{\mu_1}(\tau)}}^2+\abs{\braket{e'e}{\psi^{\mu_1}(\tau)}}^2+\abs{\braket{o'o}{\psi^{\mu_1}(\tau)}}^2\\
            &=\frac{\mu_1^2}{2}\left( \frac{\sin^2\left( \frac{E_- \tau}{2} \right)}{E_-^2} + \frac{\sin^2\left( \frac{E_+ \tau}{2} \right)}{E_+^2} \right),
        \end{split}
    \end{equation}
    This introduces a superposition of two frequencies $E_-$ and $E_+$ in the leakage frequency, where the envelope frequency is $E_-+E_+=4\Delta$ and the carrier frequency is $E_+-E_-=\Gamma$.

    \section{Leakage due to $\mu_2$}
    The leakage effect in the other onsite chemical potential is $\mu_2$ (or $\mu_3$).
    Namely, we set $\mu_1=\mu_3=\mu_4=0$, and $t=\Delta$. 
    This leads to the same noninteracting part $H^{\mu_2}_0=H^{\mu_1}_0$ as in Eq.~\eqref{eq:H_mu1}, while the perturbation term $H^{\mu_2}_1$ is 
    \begin{equation}
        H^{\mu_2}_1= \begin{pmatrix}
            H_1 & h_{\mu_2}\\
            h_{\mu_2}^\dagger & 0
        \end{pmatrix}
    \end{equation}
    where $H_1$ is in Eq.~\eqref{eq:H1} and $h_{\mu_2}$ is
    \begin{equation}
        h_{\mu_2}=\frac{1}{2}\begin{pmatrix}
            0 & 0 & -\mu_2 & 0 \\
            0 & 0 & 0 & -\mu_2\\
            -\mu_2 & 0 & 0 & 0\\
            0 & -\mu_2 & 0 & 0
        \end{pmatrix}
    \end{equation}

    Here, the noninteracting term $H^{\mu_2}_0$ has the same eigenvalues (with $\mu_1$ replaced with $\mu_2$) and eigenvectors as in Eq.~\eqref{eq:eigen}, and therefore, the final state in the Schr\"odinger picture starting from $\ket{ee}$ is
    \begin{equation}
        \psi^{\mu_2}(\tau)=
        \begin{pmatrix}
            e^{i \tau \left(2 \Delta - \frac{\mu_{2}}{2}\right)} \cos{\left(\frac{\tau \Gamma}{2} \right)}\\
            i e^{i \tau \left(2 \Delta - \frac{\mu_{2}}{2}\right)} \sin{\left(\frac{\tau \Gamma}{2} \right)}\\
            \frac{\Gamma}{4 \Delta}e^{- \frac{i \tau \mu_{2}}{2}} \sin{\left(2 \tau \Delta \right)} \sin{\left(\frac{\tau \Gamma}{2} \right)}\\
            \frac{i \Gamma }{4 \Delta}e^{- \frac{i \tau \mu_{2}}{2}} \sin{\left(2 \tau \Delta \right)} \cos{\left(\frac{\tau \Gamma}{2} \right)}\\
            \frac{\mu_{2}}{4}e^{i \tau \left(\Delta - \frac{\mu_{2}}{2}\right)}\left[ \left(\frac{i \sin{\left(\frac{E^{ }_{-} \tau}{2} \right)}}{E^{ }_{-}} - \frac{i \sin{\left(\frac{E^{'}_{+} \tau}{2} \right)}}{E^{'}_{+}}\right) e^{- \frac{i \tau \Gamma}{4}} + \left(- \frac{i \sin{\left(\frac{E^{'}_{-} \tau}{2} \right)}}{E^{'}_{-}} + \frac{i \sin{\left(\frac{E^{ }_{+} \tau}{2} \right)}}{E^{ }_{+}}\right) e^{\frac{i \tau \Gamma}{4}} \right]\\
            \frac{\mu_{2}}{4}e^{i \tau \left(\Delta - \frac{\mu_{2}}{2}\right)}\left[ \left(\frac{i \sin{\left(\frac{E^{ }_{-} \tau}{2} \right)}}{E^{ }_{-}} + \frac{i \sin{\left(\frac{E^{'}_{+} \tau}{2} \right)}}{E^{'}_{+}}\right) e^{- \frac{i \tau \Gamma}{4}} + \left(- \frac{i \sin{\left(\frac{E^{'}_{-} \tau}{2} \right)}}{E^{'}_{-}} - \frac{i \sin{\left(\frac{E^{ }_{+} \tau}{2} \right)}}{E^{ }_{+}}\right) e^{\frac{i \tau \Gamma}{4}} \right]\\
            \frac{\mu_{2}}{4}e^{i \tau \left(\Delta - \frac{\mu_{2}}{2}\right)}\left[ \left(\frac{i \sin{\left(\frac{E^{ }_{-} \tau}{2} \right)}}{E^{ }_{-}} + \frac{i \sin{\left(\frac{E^{'}_{+} \tau}{2} \right)}}{E^{'}_{+}}\right) e^{- \frac{i \tau \Gamma}{4}} + \left( \frac{i \sin{\left(\frac{E^{'}_{-} \tau}{2} \right)}}{E^{'}_{-}} + \frac{i \sin{\left(\frac{E^{ }_{+} \tau}{2} \right)}}{E^{ }_{+}}\right) e^{\frac{i \tau \Gamma}{4}} \right]\\
            \frac{\mu_{2}}{4}e^{i \tau \left(\Delta - \frac{\mu_{2}}{2}\right)}\left[ \left(-\frac{i \sin{\left(\frac{E^{ }_{-} \tau}{2} \right)}}{E^{ }_{-}} + \frac{i \sin{\left(\frac{E^{'}_{+} \tau}{2} \right)}}{E^{'}_{+}}\right) e^{- \frac{i \tau \Gamma}{4}} + \left( -\frac{i \sin{\left(\frac{E^{'}_{-} \tau}{2} \right)}}{E^{'}_{-}} + \frac{i \sin{\left(\frac{E^{ }_{+} \tau}{2} \right)}}{E^{ }_{+}}\right) e^{\frac{i \tau \Gamma}{4}} \right]
        \end{pmatrix},
    \end{equation}
    where $E_\pm=2\Delta\pm\Gamma/2$ and $E_\pm^{'}=2\Delta\pm 3\Gamma/2$.

    Therefore, the leakage to the $\ket{e'e'}$ and $\ket{o'o'}$ is the same as Eq.~\eqref{eq:P_leak_mu1}, and the leakage to $ \ket{ee'}$,  $ \ket{oo'}$, $ \ket{e'e}$, and $ \ket{o'o}$ is
    \begin{equation}\label{eq:P_leak_mu2}
        \begin{split}
            P_{ee'}(\tau)+P_{oo'}(\tau)+P_{e'e}(\tau)+P_{o'o}(\tau)&=\abs{\braket{ee'}{\psi^{\mu_2}(\tau)}}^2+\abs{\braket{oo'}{\psi^{\mu_2}(\tau)}}^2+\abs{\braket{e'e}{\psi^{\mu_2}(\tau)}}^2+\abs{\braket{o'o}{\psi^{\mu_2}(\tau)}}^2\\
            &=\frac{\mu_2^2}{4}\left( \frac{\sin^2\left( \frac{E_- \tau}{2} \right)}{E_-^2} + \frac{\sin^2\left( \frac{E_+ \tau}{2} \right)}{E_+^2} + \frac{\sin^2\left( \frac{E_-^{'} \tau}{2} \right)}{{E_-^{'}}^2} + \frac{\sin^2\left( \frac{E_+^{'} \tau}{2} \right)}{{E_+^{'}}^2} \right).
        \end{split}
    \end{equation}


    \section{Dephasing due to disorder in $\Gamma$}

    In this section, we consider the dephasing effect which is used to estimate $T_2$.
    To focus only on the low energy sector, we work in the minimal 2-level system where the Hilbert space only includes $\ket{ee}$ and $\ket{oo}$.
    The effect of $\Gamma$ acts like the magnetic field along the $x$ direction, and therefore, the effective two-level Hamiltonian is
    \begin{equation}
        H_x=B_x\sigma_x,
    \end{equation}
    where $B_x$ is quasi-static disorder following the Gaussian distribution with the variance of $\sigma_{B_x}^2$ and mean of $\bar{B}_x$, i.e., $B_x \sim \mathcal{N}(\bar{B}_x,\sigma_{B_x}^2)$, $\sigma_x$ is the Pauli X matrix, and the initial state is $\psi_x(\tau=0)=\ket{ee}$.

    Under the evolution of $H_x$, the final state $\psi_x(\tau)$ is
    \begin{equation}
        \psi_x(\tau)=e^{-iH_x\tau}\psi_x(\tau=0)=\begin{pmatrix}
            \cos(B_x\tau)\\
            -i\sin(B_x\tau)
        \end{pmatrix},
    \end{equation}  
    Therefore, the probability of finding the state in $\ket{oo}$ is 
    \begin{equation}
        P_{oo}(\tau)=\abs{\braket{oo}{\psi_x(\tau)}}^2=\sin^2(B_x\tau).
    \end{equation}
    Thus, the disorder-averaged probability is
    \begin{equation}
        \expval{P_{oo}(\tau)}_{B_x}= \frac{1}{\sqrt{2\pi}\sigma_{B_x}}\int_{-\infty}^{\infty}d B_x e^{-\frac{(B_x-\bar{B}_x)^2}{2\sigma_{B_x}^2}}\sin^2(B_x\tau)=\frac{1}{2}\left[ 1+e^{-{2\tau^2\sigma_{B_x}^2}} \cos(2\bar{B}_x\tau) \right].
    \end{equation}
    Therefore, the decay of the envelope of $\expval{P_{oo}(\tau)}_{B_x}$ follows the Gaussian decay with a prefactor of $e^{-{2\tau^2\sigma_{B_x}^2}}$, namely, $\beta=2$ in the ansatz in Eq.~\eqref{eq:Pee}  in the main text.  
    This provides a fundamental understanding of the ansatz in Eq.~\eqref{eq:Pee} in the main text.

    \section{Dephasing due to disorder in $\mu_i$, $\Delta$, and $t$}\label{app:dephasing_mu}

    Besides the dephasing effect due to the disorder in $\Gamma$, we also consider the dephasing effect due to the disorder in $\mu_i$, $\Delta$, and $t$.

    In practice, the disorder of these quantities acts like the magnetic field along the $z$ direction, and therefore, the effective two-level Hamiltonian is
    \begin{equation}
        H_{xz}=B_x\sigma_x+B_z\sigma_z,
    \end{equation}
    where $B_x$ here is constant, and $B_z$ is quasi-static disorder following the Gaussian distribution with the variance of $\sigma_{B_z}^2$ and mean of 0, i.e., $B_z \sim \mathcal{N}(0,\sigma_{B_z}^2)$, $\sigma_z$ is the Pauli Z matrix, and the initial state is again $\psi_{xz}(\tau=0)=\ket{ee}$.

    Under the evolution of $H_{xz}$, the final state $\psi_{xz}(\tau)$ is
    \begin{equation}
        \psi_{xz}(\tau)=e^{-iH_{xz}\tau}\psi_{xz}(\tau=0)=\begin{pmatrix}
            \cos(\sqrt{B_x^2+B_z^2}\tau)-i\sin(\sqrt{B_x^2+B_z^2}\tau)\cos(\theta)\\
            -i\sin(\sqrt{B_x^2+B_z^2}\tau)\sin(\theta)
        \end{pmatrix},
    \end{equation}
    where $\tan\theta=B_x/B_z$, and the probability of finding the state in $\ket{oo}$ is
    \begin{equation}
        P_{oo}(\tau)=\abs{\braket{oo}{\psi_{xz}(\tau)}}^2=\sin^2\left( \sqrt{B_x^2+B_z^2}\tau \right)\frac{B_x^2}{B_x^2+B_z^2}.
    \end{equation}
    Therefore, the disorder-averaged probability is
    \begin{equation}
        \begin{split}
            \expval{P_{oo}(\tau)}_{B_z}&= \frac{1}{\sqrt{2\pi}\sigma_{B_z}}\int_{-\infty}^{\infty}d B_z e^{-\frac{B_z^2}{2\sigma_{B_z}^2}}\sin^2\left( \sqrt{B_x^2+B_z^2}\tau \right)\frac{B_x^2}{B_x^2+B_z^2}\\
            &\approx \frac{1}{\sqrt{2\pi}\sigma_{B_z}}\int_{-\infty}^{\infty}d B_z e^{-\frac{B_z^2}{2\sigma_{B_z}^2}} \sin^2\left( \left( B_x^2+\frac{B_z^2}{2B_x} \right)\tau \right) \\
            &=\frac{1}{4}\left( 2- \frac{B_xe^{2iB_x\tau}}{\sqrt{B_x(B_x-2i\sigma_{B_z}^2\tau)}} - \frac{B_xe^{-2iB_x\tau}}{\sqrt{B_x(B_x+2i\sigma_{B_z}^2\tau)}} \right)
        \end{split}
    \end{equation}
    where the second line assumes the $B_z\ll B_x$ such that $\sqrt{B_x^2+B_z^2} \approx B_x+\frac{B_z^2}{2B_x}$, and $\frac{B_x^2}{B_x^2+B_z^2}\approx 1$.
    Therefore, it shows a power-law decay with $\expval{P_{oo}(\tau)}_{B_z} \sim \tau^{-\frac{1}{2}}$.

{
We provide the parameter for the numerical simulations of qubit dephasing in Figs.~\ref{fig:dephasing}(d-f). 
}
{
For the charge qubit in Fig.~\ref{fig:dephasing}(d), we set $\varepsilon_{L/R}\sim\mathcal{N}(0,3^2)$, $\Gamma_{L/R}\sim\mathcal{N}(5,0.05^2)$ during the $x$ pulse (i.e., with finite $\Gamma_{L/R}$) in the unit of meV, where $\mathcal{N}(\mu,\sigma^2)$ is the normal distribution with mean $\mu$ and variance $\sigma^2$; 
during the $z$ pulse (i.e., with finite $\varepsilon_{L/R}$), we set $\varepsilon_L\sim\mathcal{N}(20,3^2)$, $\varepsilon_R\sim\mathcal{N}(-20,3^2)$, $\Gamma_{L/R}\sim\mathcal{N}(0,0.05^2)$ in the unit of meV.
}
{
For the small-gap Majorana qubit (all four $\Delta_{1,2,3,4}=12$ meV) as shown in Fig.~\ref{fig:dephasing}(e), during $x$ pulse, we set four $\mu_{1,2,3,4}\sim \mathcal{N}(0,3^2)$, $t_{L/R}\sim\mathcal{N}(12,0.016^2)$ and $\Gamma=0.5$ meV; 
during $z$ pulse, we set four $\mu_{1,2,3,4}\sim \mathcal{N}(0,3^2)$, $t_{L}\sim\mathcal{N}(14,0.016^2)$, $t_{R}\sim\mathcal{N}(12,0.016^2)$ and $\Gamma=0$ meV.
}
{
For the large-gap Majorana qubit (all four $\Delta_{1,2,3,4}=38$ meV) as shown in Fig.~\ref{fig:dephasing}(f), during $x$ pulse, we set four $\mu_{1,2,3,4}\sim \mathcal{N}(0,0.3^2)$, $t_{L/R}\sim\mathcal{N}(38,0.016^2)$ and $\Gamma=0.5$ meV; during $z$ pulse, we set four $\mu_{1,2,3,4}\sim \mathcal{N}(0,0.3^2)$, $t_{L}\sim\mathcal{N}(40,0.016^2)$, $t_{R}\sim\mathcal{N}(38,0.016^2)$ and $\Gamma=0$ meV.
}


\section{Quantum capacitance measurement of a Majorana qubit}\label{app:readout}
In this section, we show that quantum capacitance measurement is capable of reading out $\ket{ee}$ and $\ket{oo}$ in a Majorana qubit.
We first review the calculations in a single Kitaev chain, and then generalize it to a Majorana qubit composed of double Kitaev chains.
The zero-temperature quantum capacitance of a state is defined as
\begin{align}
C_q = - \frac{\partial^2 E}{\partial V^2_g},
\end{align}
where $E$ is the eigenenergy of the state, and $V_g$ is the gate voltage.
The Hamiltonian of a minimal Kitaev chain can be decomposed into even- and odd-parity sectors due to fermi parity conservation. 
The even-parity Hamiltonian is
\begin{align}
H_{\text{even}} = 
\begin{pmatrix}
    \ket{00}\\
    \ket{11}
\end{pmatrix}^T
\begin{pmatrix}
0 & \Delta \\
\Delta & \mu_1 + \mu_2
\end{pmatrix}
\begin{pmatrix}
    \bra{00}\\
    \bra{11}
\end{pmatrix}, 
\end{align}
where $\mu_{1,2}$ are the onsite energies of the two dots and $t, \Delta$ are the normal and Andreev couplings.
The ground state energy $E_e$ of $\ket{e}$ is 
\begin{align}
E_e = \frac{\mu_1 + \mu_2}{2} - \sqrt{\Delta^2 + \left( \frac{\mu_1 + \mu_2}{2} \right)^2}.
\end{align}
On the other hand, the derivative with respect to gate voltage is
\begin{align}
    \frac{\partial }{\partial V_g} = \alpha_1 \frac{\partial }{\partial \mu_1} + \alpha_2 \frac{\partial }{\partial \mu_2} \approx \alpha \left( \frac{\partial }{\partial \mu_1} + \frac{\partial }{\partial \mu_2} \right),
\end{align}
where $\alpha_i \equiv d \mu_i / dV_g$ is the lever arm, and here we assume that all dots share a similar value of lever arm $\alpha$.
Therefore, it is straightforward to obtain the quantum capacitance of the even-parity ground state as below
\begin{align}
    C^e_{q} =- \frac{\partial^2 E_e}{\partial V^2_g} = \frac{\alpha^2}{ \sqrt{\Delta^2 + \left( \frac{\mu_1 + \mu_2}{2} \right)^2} } 
    -  \frac{\alpha^2 (\mu_1 + \mu_2)^2}{4 \left( \Delta^2 + \left( \frac{\mu_1 + \mu_2}{2} \right)^2 \right)^{3/2}}.
\end{align}
At the sweet spot of $\mu_1 = \mu_2=0$, we have
\begin{align}
    C^e_{q} = \frac{\alpha^2}{ \Delta }.
\end{align}
On the other hand, the Hamiltonian in the odd-parity sector is
\begin{align}
H_{\text{odd}} = 
\begin{pmatrix}
    \ket{10}\\
    \ket{01}
\end{pmatrix}^T
\begin{pmatrix}
\mu_1 & t \\
t &  \mu_2
\end{pmatrix}
\begin{pmatrix}
    \bra{10}\\
    \bra{01}
\end{pmatrix}, 
\end{align}
with ground-state energy being
\begin{align}
E_o = \frac{\mu_1 + \mu_2}{2} - \sqrt{t^2 + \left( \frac{\mu_1 - \mu_2}{2} \right)^2}.
\end{align}
The corresponding quantum capacitance is
\begin{align}
    C^o_{q} = 0,
\end{align}
due to the opposite signs of coefficients in front of $\mu_1$ and $\mu_2$ in the term of $(\frac{\mu_1 - \mu_2}{2})^2$.

We now generalize our calculations to the quantum capacitance of the double Kitaev chain system.
The eigenenergies of the ground states are simply the sum of the left and right chains, i.e.,
\begin{align}
    E_{ab} = E_{aL}(\mu_{L1},\mu_{L2} ) + E_{bR}(\mu_{R1},\mu_{R2} ).
\end{align}
where $a,b$ denotes the parity $e,o$.
We note that the $\mu$ dependence of $E$ is separable between the left and right chains, which indicates that $\partial/\partial V_g \to \alpha \left( \frac{\partial }{\partial \mu_{L1}} + \frac{\partial }{\partial \mu_{L2}} \right)$ for the left chain energy, while $\partial/\partial V_g \to \alpha \left( \frac{\partial }{\partial \mu_{R1}} + \frac{\partial }{\partial \mu_{R2}} \right)$ for the right one.
Therefore, we have 
\begin{align}
C^{ab}_{q} = - \frac{\partial^2 E_{ab}}{\partial V^2_g} = - \frac{\partial^2 E_{aL}}{\partial V^2_g}- \frac{\partial^2 E_{bR}}{\partial V^2_g}
= C^{aL}_{q} + C^{bR}_{q},
\end{align}
that is, the quantum capacitance of the state in the whole system is a sum of the value in each chain separately.
We therefore have
\begin{align}
& C^{ee}_{q} = \frac{\alpha^2}{ \Delta_L } + \frac{\alpha^2}{ \Delta_R }, \nonumber \\
& C^{oo}_{q} = 0, \nonumber \\
& C^{eo}_{q} = \frac{\alpha^2}{ \Delta_L } , \nonumber \\
& C^{oe}_{q} = \frac{\alpha^2}{ \Delta_R }.
\end{align}
Therefore, one can distinguish between $\ket{ee}$ and $\ket{oo}$ states using quantum capacitance measurement.
Furthermore, the values of $C_q$ for $\ket{eo}$ and $\ket{oe}$ are generally very different from the qubit states.
Thus our method also provides a possible way to investigate quasiparticle poisoning effect by analyzing the readout results of $\ket{eo}$ and $\ket{oe}$.

\end{document}